\documentclass[onecolumn,aps,preprint,final,amsmath,amssymb,groupedaddress]{revtex4-1}
\usepackage{times}
\usepackage{graphicx}
\usepackage{amssymb}
\usepackage{amsthm}
\usepackage{amsmath}
\usepackage{dsfont}
\usepackage{bm}
\usepackage{bbm}
\usepackage{mathrsfs}
\usepackage{bbold}
\usepackage{color}

\begin{document}
 
\title{
Casimir self-stress in a dielectric sphere
}
\author{Yael Avni}
\author{Ulf Leonhardt}
\affiliation{Department of Physics of Complex Systems,
Weizmann Institute of Science, Rehovot 76100, Israel}

\date{\today}

\begin{abstract}
The dielectric sphere has been an important test case for understanding and calculating the vacuum force of a dielectric body onto itself. Here we develop a method for computing this force in homogeneous spheres of arbitrary dielectric properties embedded in arbitrary homogeneous backgrounds, assuming only that both materials are isotropic and dispersionless. Our results agree with known special cases; most notably we reproduce the prediction of Boyer and Schwinger et al.\ of a repulsive Casimir force of a perfectly reflecting shell. Our results disagree with the literature in the dilute limit. We argue that Casimir forces can not be regarded as due to pair-wise Casimir-Polder interactions, but rather due to reflections of virtual electromagnetic waves.\\

\end{abstract}
 
\maketitle

\section{Introduction}
\begin{figure}[h!]
\begin{center}
\includegraphics[width=6cm]{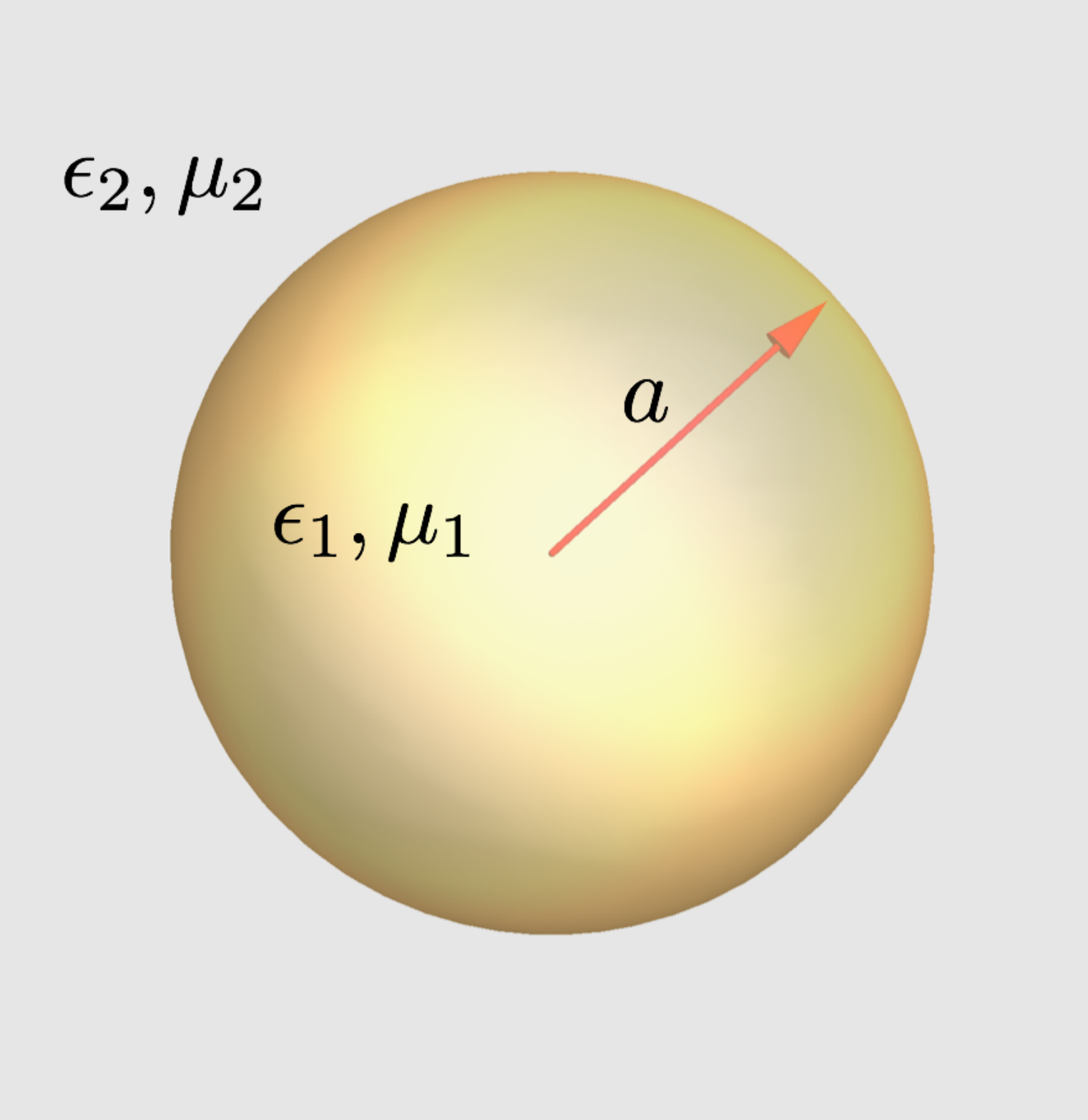}
\caption{
{\small
Dielectric sphere of radius $a$ surrounded by dielectric background. The sphere is subject to pressure arising from vacuum fluctuations. The dielectric properties of the sphere and the background are incorporated in their permittivity and permeability ($\epsilon$ and $\mu$ respectively), which determine the response of the materials to vacuum fluctuations.}
}
\label{sphere}
\end{center}
\end{figure}

Julian Schwinger {\it et al.}, in a paper from 1977 \cite{VanEnk1995}, described the Casimir force as ``one of the least intuitive consequences of quantum electrodynamics''. Four decades later, although extensive work has been made in the field, the mystery has yet to disappear. Observed to good accuracy in experiments \cite{Lamoreaux1997, Mohideen1998, decca2015measuring} , Casimir forces are known to originate from the ubiquitous vacuum fluctuations of the electromagnetic field that excite dielectric and conducting materials \cite{Buhmann,Philbin2010}, causing them to interact with one another \cite{Scheel2015}. Our ability of predicting these forces, however, is limited, in particular for predictions of Casimir forces inside dielectric bodies \cite{Philbin,Simpson,Horsley,SimpsonSurprise,griniasty2017casimir}. Inspired by the work \cite{Milton1978} of Milton, DeRaad and Schwinger on the Casimir self force of a perfectly conducting shell, we solve the problem of calculating the Casimir stress in an arbitrary dielectric sphere embedded in an arbitrary dielectric background, both assumed to be isotropic and dispersionless (Fig.~\ref{sphere}). Our findings could be experimentally tested by probing the resonances of microspheres. Our results agree with known special cases \cite{Brevik1982,Milton1997,Brevik1998}, but disagree with what was perceived as the dilute limit \cite{Brevik1994, Brevik1999, Barton}. We found that the physical picture of Casimir forces as arising due to pair-wise van der Waals interactions is no longer justified in the dilute limit of the dielectric sphere. We develop an alternative physical picture based on reflections of virtual waves instead of summations of forces. Such a change of perspective may have implications for a wide range of Casimir--related phenomena. 

Thanks to an immense progress in numerical methods over the last decade \cite{Reid}, the Casimir force between any arbitrary number of distinct objects with finite separation between them can be resolved using path integral approach combined with scattering theory (also known as the EGJK method) \cite{Rahi-2009, Rodriguez2011}. However, this method relies on renormalizing the force by subtracting the scattering of each individual object when it is infinitely separated from the rest, which cannot be done when considering the Casimir self stress that each object experiences upon itself. Therefore this method cannot predict the self stress of a dielectric sphere, which is the main focus of this paper.

\begin{figure}[h!]
\includegraphics[width=12cm]{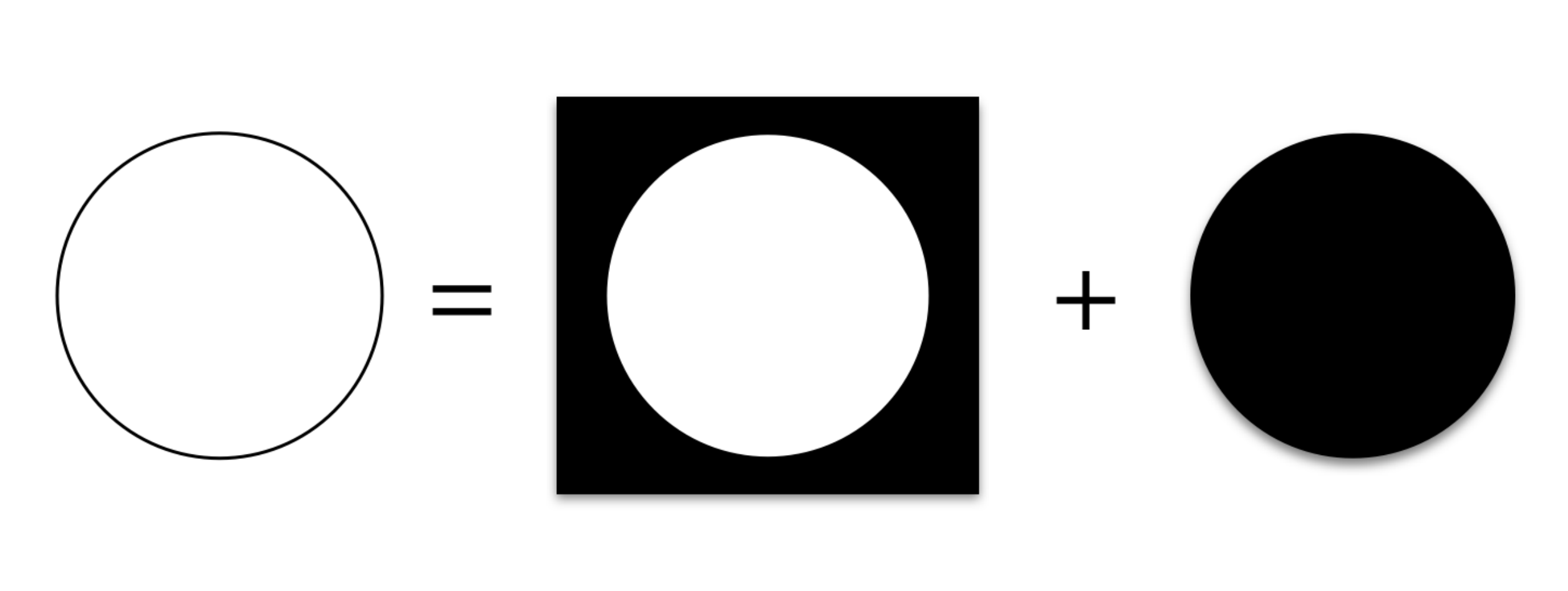}\centering
\caption{{\small A conducting spherical shell can be viewed as the sum over two cases: a vacuum cavity in a conducting background and a conducting sphere in vacuum.}}
\label{sum}
\end{figure}

The spherical problem dates back to Casimir himself, who proposed that vacuum fluctuations might cause a conducting spherical shell to attract itself, in a way analogous to the case of two conducting plates \cite{Casimir1948a, Casimir1953}. The problem of the spherical shell can be reduced to the problem of the sphere (Fig.~\ref{sum}). Casimir's motivation came from a semi-classical model for the structure of an electron in which the Casimir stress might have played an important roll (similarly, the Casimir stress was considered to affect the ``bag model", a model of hadrons \cite{Bagmodel, bagMilton1980}). It was rather a surprise when Boyer \cite{Boyer1968} showed in 1968 that the Casimir stress in a perfectly conducting spherical shell is repulsive, {\it i.e.}\ tends to expand the sphere \cite{Boyer1968} (although two conducting hemispheres attract each other \cite{Kenneth2006}). Boyer's calculation, which was based on a mode summation method, lacked a physical justification for some of its renormalization techniques,  but his result was re-derived by different papers \cite{Balian1977,Nesterenko1998} including the work of Milton, DeRaad and Schwinger \cite{Milton1978} on which the present paper has greatly relied on.

Since then, there have been several attempts to generalize this result from the perfectly conducting spherical shell to dielectric spheres, most notably the pioneering work by Milton \cite{Milton1980}. They all found that the Casimir force of a dielectric sphere is cutoff dependent, and no clear results were obtained, except for two special cases: the dilute ball \cite{Brevik1994, Brevik1999, Barton, Bordag1999}, and the case where the speed of light is identical both inside and outside the sphere (but $\epsilon$ and $\mu$, the permittivity and permeability, are different) \cite{Brevik1982,Milton1997, Brevik1998}. In these special cases a finite term, independent of any cutoff was extracted and the additional diverging terms were sometimes dismissed either by a mathematical trick or simply by claiming that these terms are unobservable. To our knowledge, there is no clear statement regarding the physicality of the cutoff dependent terms.
To quote from a thorough review of developments in the Casimir effect published in Ref.\ \cite{Bordag20011}:

\textit{``Despite the mentioned results the situation with a dielectric body remains unsatisfactory... It is impossible
to identify a unique quantum energy. But on the other hand, we are confronted with real macroscopic bodies and the clear existence of vacuum fluctuations of the electromagnetic field constituting a real physical situation, so that no infinities or arbitrariness should occur"}.

In this paper we present a method to calculate the Casimir self stress of a homogeneous sphere inside a homogeneous background, which is a modification of previous calculations: it uses the distance from the radius of the sphere as a regularizer rather than a frequency or a wavenumber cutoff. Our method is inspired from correcting an unjustified mathematical procedure we spotted in previous calculations, namely the interchange between a limit and a sum which should be made the other way round. Taking the correct order of limits we develop a way of extracting the \textit{finite macroscopic contribution} to the Casimir self stress for general dielectric constants (but constants in frequency). We interpret the other terms as originating from the microscopic description, that includes the finite separation between the atoms in a medium, which is lost in the macroscopic description. Our method applies only to dispersion-less materials since, to our understanding, only in this case there is a clear distinction between the finite macroscopic term and the other cutoff dependent terms. We calculate the macroscopic contribution to the self stress as a function of the dielectric constant and give an estimation for the correction to the surface tension of the sphere that arises from it, which scales like $a^{-3}$, $a$ being the radius of the sphere. We show that this method, although different in essence, reproduces known results such as the stress in the limit of a perfectly conducting spherical shell and the case in which the speed of light is identical both inside and outside the sphere. Although the finite terms are reproduced, we interpret them differently: they are the macroscopic contributions to the stress, but microscopic corrections exist, as is evident from the cutoff dependent terms. Our results disagree with the previous calculation of the dilute case (a sphere with $\epsilon \approx 1$ in vacuum) \cite{Brevik1999}. We show that by taking the correct order of limits an additional term which is linear in $\epsilon-1$ emerges. This suggests that the Casimir self stress cannot be understood by a simple pairwise summation over van der Waals interactions as claimed in Ref.~\cite{Brevik1999}. We present an alternative picture based on reflections of the electromagnetic field from the boundaries. This picture coincides with the van der Waals interactions picture when the boundaries are flat, as in a system of two conducting plates.

The paper is organized in the following way: In Sec.~\ref{introA} we briefly review the framework of Lifshitz theory. In Sec.~\ref{formulae} we develop our method of calculating the stress by keeping a finite distance from the surface of the sphere and derive from it a formula for the macroscopic part of the self-stress in a dielectric sphere. In Sec.~\ref{numeric} we show numerical results for this contribution to the stress for different cases, compare them to previous results and discuss their meaning. In Sec.~\ref{dispersion} we discuss the inclusion of dispersion.

\section{Renormalization and Lifshitz theory} \label{introA}

The main tool in this work is Lifshitz theory \cite{LifshitzE.M.1956, Dzyaloshinskii1961} (and its interpretation in Refs.\ \cite{Scheel2015,Leonhardt2010}) that supplies a comprehensive framework to calculate the Casimir force in dielectric materials. In this theory, the force density,
\begin{equation} \label{force density1}
\bm{f}=\nabla\cdot\sigma, 
\end{equation}
is described in terms of the vacuum expectation value of Maxwell's electromagnetic stress tensor:
\begin{equation}
\sigma=\langle0|\hat{\sigma}|0\rangle=\langle0|\hat{\bm{E}}\otimes\hat{\bm{D}}|0\rangle+\langle0|\hat{\bm{B}}\otimes\hat{\bm{H}}|0\rangle-\frac{1}{2}\bigg(\langle0|\hat{\bm{E}}\cdot\hat{\bm{D}}|0\rangle+\langle0|\hat{\bm{B}}\cdot\hat{\bm{H}}|0\rangle\bigg)\mathbbm{1}_{3} \,.
\label{stress}
\end{equation}
Here \(\mathbbm{1}_3\) denotes the three-dimensional identity matrix and $\hat{\bm{D}}$ and $\hat{\bm{H}}$ are related to $\hat{\bm{E}}$ and $\hat{\bm{B}}$ by the constitutive equations:
\begin{equation}
\hat{\bm{D}}=\epsilon_{0}\epsilon\hat{\bm{E}},\,\,\,\,\hat{\bm{B}}=\epsilon_{0}\epsilon\hat{\bm{H}},\,\,\,\,\epsilon_{0}\mu_{0}=c^{-2}.
\end{equation}

A naive calculation of the Casimir force produces infinity, which is due to the divergence of elements of the form:  \(\langle0|\hat{\bm{E}}(\textbf{r})^{2}|0\rangle,\langle0|\hat{\bm{B}}(\textbf{r})^{2}|0\rangle\). However, these infinities are usually not physical. To see this we regularize the divergence by ``point splitting'':
\begin{equation}
\begin{gathered}
\langle0|\hat{\bm{E}}(\textbf{x})\otimes\hat{\bm{E}}(\textbf{x})|0\rangle\to\langle0|\hat{\bm{E}}(\textbf{x})\otimes\hat{\bm{E}}(\textbf{x}')|0\rangle
\\
\langle0|\hat{\bm{B}}(\textbf{x})\otimes\hat{\bm{B}}(\textbf{x})|0\rangle\to\langle0|\hat{\bm{B}}(\textbf{x})\otimes\hat{\bm{B}}(\textbf{x}')|0\rangle
\end{gathered}
\end{equation}
where \(\textbf{x}\) is a 4-vector. As long as $\textbf{x}\neq\textbf{x}'$ the expectation values are finite, and so the limit \(\textbf{x} \to \textbf{x}' \) it taken only at the very end.

The above correlators can be obtained from a single fundamental function: the classical Green's function \(G\), which is defined by the equation:
\begin{equation} \label{green function eq}
\nabla\times\frac{1}{\mu}\nabla\times G-\epsilon\frac{\omega^{2}}{c^{2}}G=\mathbbm{1}_{3}\,\delta\left(\bm{r}-\bm{r}'\right).
\end{equation}
The Green's function \(G\) is a second rank tensor that is proportional to the classical electric field at spatial position \(\bm{r}\) generated by a dipole at spatial position \(\bm{r}'\) oscillating with frequency \(\omega\). The dipole may point in all three spatial directions, which is described by the unity matrix in the right-hand side of Eq.\ (\ref{green function eq}). The relations between the field correlators and \(G\) follow from the quantum theory of electromagnetism in media \cite{Buhmann,Philbin2010,Scheel2015} and in particular from the fluctuation-dissipation theorem \cite{Scheel2015}:
\begin{equation}
\langle0|\widetilde{\bm{A}}\left(\bm{r},\omega\right)\otimes\widetilde{\bm{A}}\left(\bm{r}',\omega'\right)|0\rangle = -\frac{\hbar}{\varepsilon_0 c^2\pi}\,(2\pi)^2\, \mathrm{Im}\, G\,\delta(\omega-\omega')
\label{theorem}
\end{equation}
for the vacuum correlator of the Fourier-transformed vector potential $\hat{\bm{A}}$ at positive frequencies (the correlator vanishes for negative frequencies). In fact, one obtains \cite{Scheel2015} from $\bm{E}=-\partial\bm{A}/\partial t$ and $\bm{B}=\nabla\times\bm{A}$ by Fourier transformation in the limit $t'\rightarrow t$:
\begin{equation} \label{correlatorsim}
\begin{gathered}\langle0|\hat{\bm{E}}\left(\bm{r}\right)\otimes\hat{\bm{E}}\left(\bm{r}'\right)|0\rangle=-\frac{\hbar}{\pi\epsilon_{0}c^{2}}\int\limits _{0}^{\infty}\xi^{2}G\left(\bm{r},\bm{r}',i\xi\right)\text{d}\xi\,,\\
\langle0|\hat{\bm{B}}\left(\bm{r}\right)\otimes\hat{\bm{B}}\left(\bm{r}'\right)|0\rangle=\frac{\hbar}{\pi\epsilon_{0}c^{2}}\int\limits _{0}^{\infty}\nabla\times G\left(\bm{r},\bm{r}',i\xi\right)\times\overleftarrow{\nabla}\text{d}\xi \,.
\end{gathered}
\end{equation}
Here the property $G(-\omega)=G(\omega)^*$ was used for real frequencies $\omega$ and then the integration contours were deformed to the positive imaginary frequency axis with $\omega=\mathrm{i}\xi$ (for details see Ref.~\cite{Scheel2015}).

The use of the Green's function formalism translates the problem into the problem of finding the interaction between a fluctuating dipole at \(\bm{r}'\) and a test particle at \(\bm{r}\). The interaction between the two points can be divided into direct interaction mediated by ``outgoing waves'', and indirect interaction which consists of waves that scatter from their surroundings \cite{griniasty2017casimir}. In Lifshitz theory one subtracts the contribution of the direct interaction, which is the free Green's function (without boundary conditions except going to zero at infinity). It corresponds to the interaction of the dipole with itself which is unphysical. We are then left with the physical renormalized stress
\begin{equation}
\sigma\left(\bm{r}\right)=\lim_{\bm{r}'\to \bm{r}}\big(\sigma\left(\bm{r},\bm{r}'\right)-\sigma_{\infty}\left(\bm{r},\bm{r}'\right)\big)
\end{equation}
where $\sigma_{\infty}\left(\bm{r},\bm{r}'\right)$ is the direct interaction part of the stress.

Lifshitz theory predicts a finite Casimir force between separated materials. However, when considering a single object such as a sphere, we are interested in \textit{Casimir self stresses}. In that case cutoff dependent terms can originate from the breaking of the continuum picture, but as we will show, it will be along-side finite terms that arise solely from the macroscopic description of the problem.

\section{The force on a homogeneous sphere} \label{formulae}
Consider a homogeneous sphere of radius $a$ surrounded by a homogeneous background, with different isotropic permittivities and permeabilities \(\epsilon_{1,2},\mu_{1,2}\) at zero temperature.
We follow Ref.~\cite{Milton1978} (which contains a similar calculation for a perfectly conducting spherical shell) and merely generalize it. 

\subsection{The stress and the force density}

First we find the Green's function and from it obtain the renormalized stress components (the full derivation is described in Appendix \ref{app A}):
\begin{equation} \label{diag}
\protect\sigma=\text{diag}\left(\sigma_{r}^{r},\sigma_{\theta}^{\theta},\sigma_{\phi}^{\phi}\right)
\end{equation}
\begin{equation} \label{sigmar}
\sigma_{r}^{r}\left(r\right)=\frac{\hbar c}{8\pi^{2}r^{2}a^{2}}\sum_{l}\int\limits _{0}^{\infty}\left(2l+1\right)\left(l\left(l+1\right)+y_{i}^{2}\rho^{2}-\frac{\text{d}}{\text{d}\rho}\rho\frac{\text{d}}{\text{d}\rho'}\rho'\right)g_{l}\left(y,\rho,\rho'\right)|_{\rho'\to\rho}\,\text{d}y
\end{equation}
\begin{equation} \label{sigmatheta}
\sigma_{\theta}^{\theta}\left(r\right)=\sigma_{\phi}^{\phi}\left(r\right)=-\frac{\hbar c}{8\pi^{2}r^{2}a^{2}}\sum_{l}\int\limits _{0}^{\infty}\left(2l+1\right)l\left(l+1\right)g_{l}\left(y,\rho,\rho'\right)|_{\rho'\to\rho}\,\text{d}y
\end{equation}
where $\rho=r/a$, $y_{1}=y\sqrt{\epsilon_{1}\mu_{1}}$, $y_{2}=y\sqrt{\epsilon_{2}\mu_{2}}$, $y$ is a normalized imaginary frequency and $g_{l}$ is the function
\begin{equation}
g_{l}\left(y,\rho,\rho'\right)=\begin{cases}
C_{\text{i}}\sqrt{\frac{1}{\rho\rho'}}I_{l+\frac{1}{2}}\left(\rho y_{1}\right)I_{l+\frac{1}{2}}\left(\rho'y_{1}\right) & r,r'<a\\
C_{\text{o}}\sqrt{\frac{1}{\rho\rho'}}K_{l+\frac{1}{2}}\left(\rho y_{2}\right)K_{l+\frac{1}{2}}\left(\rho'y_{2}\right) & r,r'>a
\end{cases}
\end{equation}
with coefficients
\begin{equation} \label{coefficients} 
\begin{gathered}C_{\text{i}}=\sum_{\nu=\epsilon,\mu}\frac{\frac{\nu_{1}}{\sqrt{y_{1}}}K_{l+\frac{1}{2}}\left(y_{1}\right)\left(\sqrt{y_{2}}K_{l+\frac{1}{2}}\left(y_{2}\right)\right)'-\frac{\nu_{2}}{\sqrt{y_{2}}}K_{l+\frac{1}{2}}\left(y_{2}\right)\left(\sqrt{y_{1}}K_{l+\frac{1}{2}}\left(y_{1}\right)\right)'}{\frac{\nu_{2}}{\sqrt{y_{2}}}K_{l+\frac{1}{2}}\left(y_{2}\right)\left(\sqrt{y_{1}}I_{l+\frac{1}{2}}\left(y_{1}\right)\right)'-\frac{\nu_{1}}{\sqrt{y_{1}}}I_{l+\frac{1}{2}}\left(y_{1}\right)\left(\sqrt{y_{2}}K_{l+\frac{1}{2}}\left(y_{2}\right)\right)'}\\
C_{\text{o}}=\sum_{\nu=\epsilon,\mu}\frac{\frac{\nu_{1}}{\sqrt{y_{1}}}I_{l+\frac{1}{2}}\left(y_{1}\right)\left(\sqrt{y_{2}}I_{l+\frac{1}{2}}\left(y_{2}\right)\right)'-\frac{\nu_{2}}{\sqrt{y_{2}}}I_{l+\frac{1}{2}}\left(y_{2}\right)\left(\sqrt{y_{1}}I_{l+\frac{1}{2}}\left(y_{1}\right)\right)'}{\frac{\nu_{2}}{\sqrt{y_{2}}}K_{l+\frac{1}{2}}\left(y_{2}\right)\left(\sqrt{y_{1}}I_{l+\frac{1}{2}}\left(y_{1}\right)\right)'-\frac{\nu_{1}}{\sqrt{y_{1}}}I_{l+\frac{1}{2}}\left(y_{1}\right)\left(\sqrt{y_{2}}K_{l+\frac{1}{2}}\left(y_{2}\right)\right)'}
\end{gathered}.
\end{equation}
The limit of a perfectly conducting spherical shell is obtained by the substitutions:
\begin{equation}
\begin{gathered}C_{\text{i}} \to -\frac{K_{l+\frac{1}{2}}\left(y\right)}{I_{l+\frac{1}{2}}\left(y\right)}-\frac{\left(\sqrt{y}K_{l+\frac{1}{2}}\left(y\right)\right)'}{\left(\sqrt{y}I_{l+\frac{1}{2}}\left(y\right)\right)'}\\
C_{\text{o}} \to -\frac{I_{l+\frac{1}{2}}\left(y\right)}{K_{l+\frac{1}{2}}\left(y\right)}-\frac{\left(\sqrt{y}I_{l+\frac{1}{2}}\left(y\right)\right)'}{\left(\sqrt{y}K_{l+\frac{1}{2}}\left(y\right)\right)'}\end{gathered}
\end{equation}
and \(\epsilon_1=\mu_1=\epsilon_2=\mu_2=1\).

Note that a similar derivation with dielectrics was done in Ref. \cite{Milton1997}, but there the focus was on the stress on the surface of the sphere, whereas we are interested in the stress as a function of the distance from the sphere (for a reason that will soon become clear).  Our results agree with Ref. \cite{Milton1997} when substituting $r=a$.

From Eq. (\ref{force density1}) we get:
\begin{equation} \label{fr}
f_{r}=\left(\nabla\cdot\sigma\right)_{r}=\frac{1}{r^{2}}\frac{\text{d}}{\text{d}r}\left(r^{2}\sigma_{r}^{r}\right)-\frac{1}{r}\sigma_{\theta}^{\theta}-\frac{1}{r}\sigma_{\phi}^{\phi}
\end{equation}
where $f_\theta=f_\phi=0$ from symmetry considerations.
By substituting the stress components in Eq. (\ref{fr}) we obtain that the force density is strictly zero whenever \(r\neq a\). Moreover, the cancelation is valid for each $l$ and $y$ in the sum and integral separately. The Casimir force differs from zero only at the ``jump", as expected.

\subsection{The force on the surface of the sphere}
Since the force density is zero anywhere but at $r=a$ and has spherical symmetry, we can write:
\begin{equation} \label{delta}
4\pi r^{2}f_{r}=4\pi r^{2}\left(\frac{1}{r^{2}}\frac{\text{d}}{\text{d}r}\left(r^{2}\sigma_{r}^{r}\right)-\frac{2}{r}\sigma_{\theta}^{\theta}\right)=F\delta\left(r-a\right).
\end{equation}
 Note that $F$ is defined as the integral of the pressure over the surface of the sphere and  not as the net force applied on the center of mass (which is of course zero).

To better understand the left hand side we analyze $r^{2}\sigma_{r}^{r}$. In Appendix \ref{app B} we show that it can be written as the following series expansion:
\begin{equation} \label{expansion}
r^{2}\sigma_{r}^{r}(\Delta)=\begin{cases}
\sum_{n=-3}^{\infty}a_{n}^{\mbox{i}}\Delta^{n}+\sum_{n=0}^{\infty}b_{n}^{\mbox{\ensuremath{\text{i}}}}\Delta^{n}\log\left(\Delta\right) & \mbox{for}\quad r<0\\
\sum_{n=-3}^{\infty}a_{n}^{\mbox{o}}\Delta^{n}+\sum_{n=0}^{\infty}b_{n}^{\mbox{\ensuremath{\text{o}}}}\Delta^{n}\log\left(\Delta\right) & \mbox{for} \quad r>0
\end{cases}
\end{equation}
where $\Delta \propto |r-a|$.

To satisfy Eq.\ (\ref{delta}), each term in the above expansion must be canceled by a counter term in the expansion of $\sigma_{\theta}^{\theta}(\Delta)$, except at the surface of the sphere. Therefore Eq.\ (\ref{delta}) becomes

\begin{equation}
4\pi\cdot\lim_{\Delta\to0}\left[\sum_{n=-3}^{\infty}\left(a_{n}^{\mbox{o}}-a_{n}^{\mbox{i}}\right)\Delta^{n}+\sum_{n=0}^{\infty}\left(b_{n}^{\mbox{o}}-b_{n}^{\mbox{i}}\right)\Delta^{n}\log\left(\Delta\right)\right]\delta\left(r-a\right)=F\delta\left(r-a\right) \,,
\end{equation}
which amounts to
\begin{equation} \label{Force full}
F=4 \pi \cdot \lim_{\Delta\to0}\left[\frac{a_{-3}^{\mbox{o}}-a_{-3}^{\mbox{i}}}{\Delta^{3}}+\frac{a_{-2}^{\mbox{o}}-a_{-2}^{\mbox{i}}}{\Delta^{2}}+\frac{a_{-1}^{\mbox{o}}-a_{-1}^{\mbox{i}}}{\Delta}+\left(b_{0}^{\mbox{o}}-b_{0}^{\mbox{i}}\right)\log\left(\Delta\right)\right]+4\pi\left(a_{0}^{\mbox{o}}-a_{0}^{\mbox{i}}\right) \,.
\end{equation}

Equation (\ref{Force full}) reveals that the force has two contributions. One is the sum of all the diverging terms (with $\Delta^{-3}$, $\Delta^{-2}$, $\Delta^{-1}$ and $\log\Delta$). These divergences result from the artificial smoothening of the medium that occurs in a macroscopic model, {\it i.e.}\ the assumption that the constituents of the medium are infinitesimally close to one another. These terms will be modified depending on the microscopic structure of the sphere and the finite separation between atoms and molecules, and therefore they are cutoff dependent. The second is the macroscopic contribution which does not require any cutoff. To calculate it, one needs to extract the finite parts in the expansion of Eq. (\ref{expansion}) and compute:
\begin{equation} \label{final force}
F_m=4 \pi (a_{0}^{\text{o}}-a_{0}^{\text{i}})
\end{equation}
(the subscript $m$ stands for macroscopic).

Looking at Eq.\ (\ref{Force full}), one should ask whether the expansion is unique, whether the constant part $a_{0}^{\mbox{o}}-a_{0}^{\mbox{i}}$ is well defined. If we substitute $\Delta\to C \Delta$ the terms which are power laws transform like $\frac{a_{n}}{\Delta^{n}}\to\frac{C^{n}a_{n}}{\left(C\Delta\right)^{n}}$ and do not affect the constants. The logarithmic term however, induces a constant as $A\mbox{log}\left(\Delta\right)\to A\mbox{log}\left(C\Delta\right)-A\mbox{log}\left(C\right)$. Therefore $a_{0}^{\mbox{o}}-a_{0}^{\mbox{i}}$ depends on the specific scale we choose. As there is only one scale in this problem --- the radius of the sphere --- it is natural to use $\delta \equiv |r-a|/a$ as the parameter of the expansion. Strictly speaking, there is no completely water-tight justification of preferring $\delta$ over $\frac{1}{2} \delta$ (the radius rather than the diameter). However, for not too large dielectric constants our results do not change by replacing $\delta$ by $C \delta$ where $C$ is of order 1, as we show in Appendix \ref{app C}.

In previous calculations \cite{Milton1980,Milton1997,Brevik1994} that used the stress-tensor approach, a finite part was extracted in a different manner: the limit of $r \to a$ was taken for each wavenumber ($l$) separately, and the summation over the wavenumber was carried out at the very end. Then, a term which did not depend on a wavenumber cutoff was obtained.

However, from Eq.\ (\ref{fr}) it is clear that the correct order of operations is {\it first} the summation, which simply gives the stress as a function of $r$. and {\it then} the limit $r \to a$. Indeed, the methods share similarities and as we will see, in some cases they produce the same results, but in others they disagree.

In the following section we compute Eq.\ (\ref{final force}) --- the macroscopic contribution to the stress --- in different cases.
\section{Numerical results and discussion} \label{numeric}
To extract the constants $a_{0}^{\text{i}}$ and $a_{0}^{\text{o}}$ we evaluated $\Delta r^2\sigma_{r}^{r}\left(r\right) \equiv r_{\text{out}}^2\sigma^r_{r,\text{out}}-r_{\text{in}}^2\sigma^r_{r,\text{in}}$ as a function of $\delta=\frac{|r-a|}{a}$ at the points $\frac{1}{100} \leq \delta \leq \frac{1}{1000}$, $\Delta\delta=\frac{1}{5000}$, and fitted it to the model:
\begin{equation} \label{fit2}
y=\sum_{n=-3}^{4}a_{n}^{\mbox{i}}\delta^{n}+\sum_{n=0}^{4}b_{n}^{\mbox{\ensuremath{\text{i}}}}\delta^{n}\log\left(\delta\right),
\end{equation}
in accordance with Eq. (\ref{expansion}) (the choice of $n=4$ as the upper bound is explained in Appendix \ref{app C}).

\begin{figure}[h!]
\includegraphics[width=14cm]{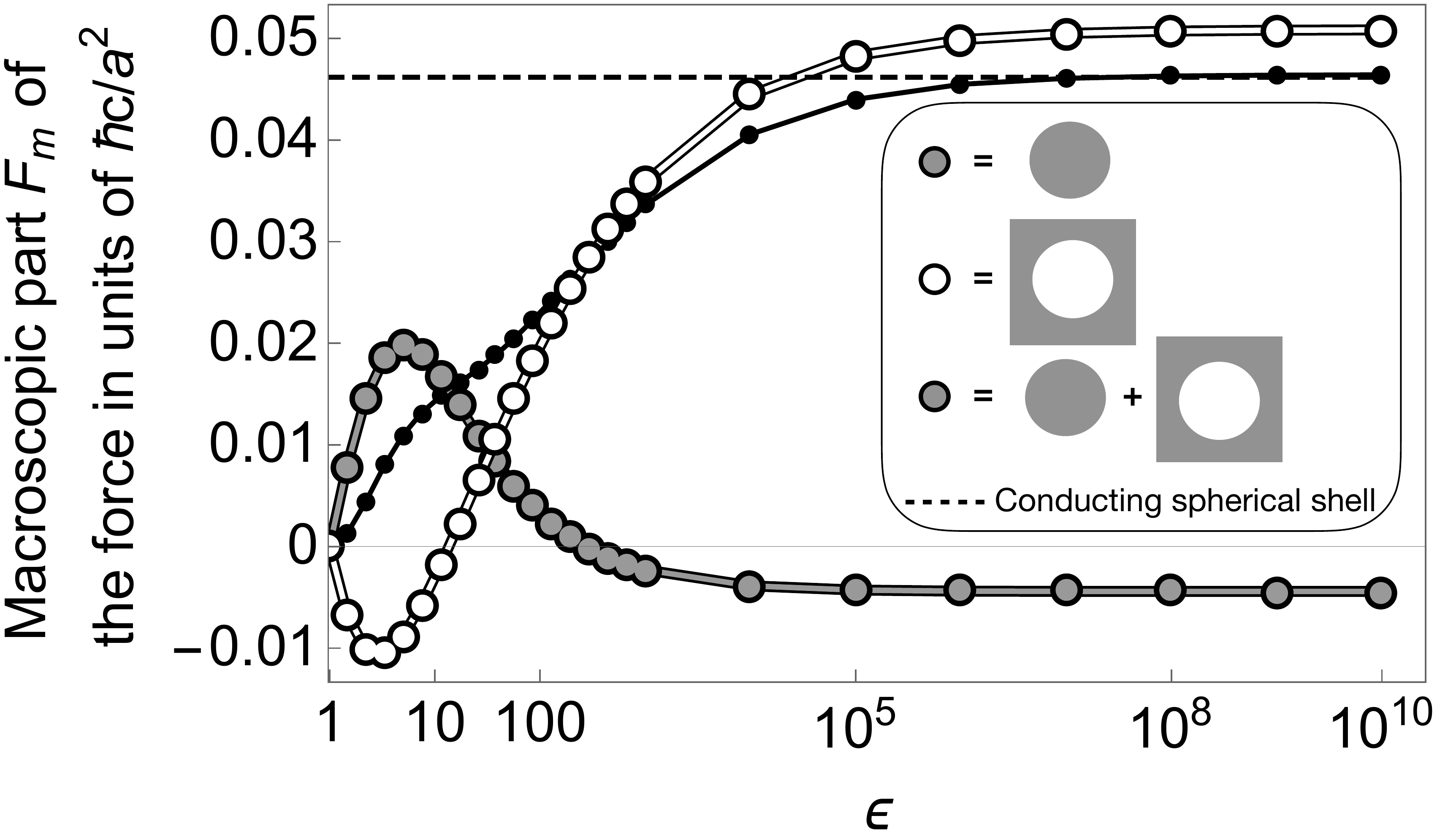}\centering
\caption {\small{$F_m$ as a function of $\epsilon$. Gray: $\epsilon_2=\mu_{1,2}=1$, $\epsilon_1=\epsilon$. White: $\epsilon_1=\mu_{1,2}=1$, $\epsilon_2=\epsilon$. Black: The sum of the gray and white points, which is meaningful only in the limit $\epsilon \to \infty$. Dashed: the limit of a perfectly conducting spherical shell: $F_m=0.04618 \hbar c/a^2$.}} \label{force}
\end{figure}

\begin{figure}[h!]
\includegraphics[width=10cm]{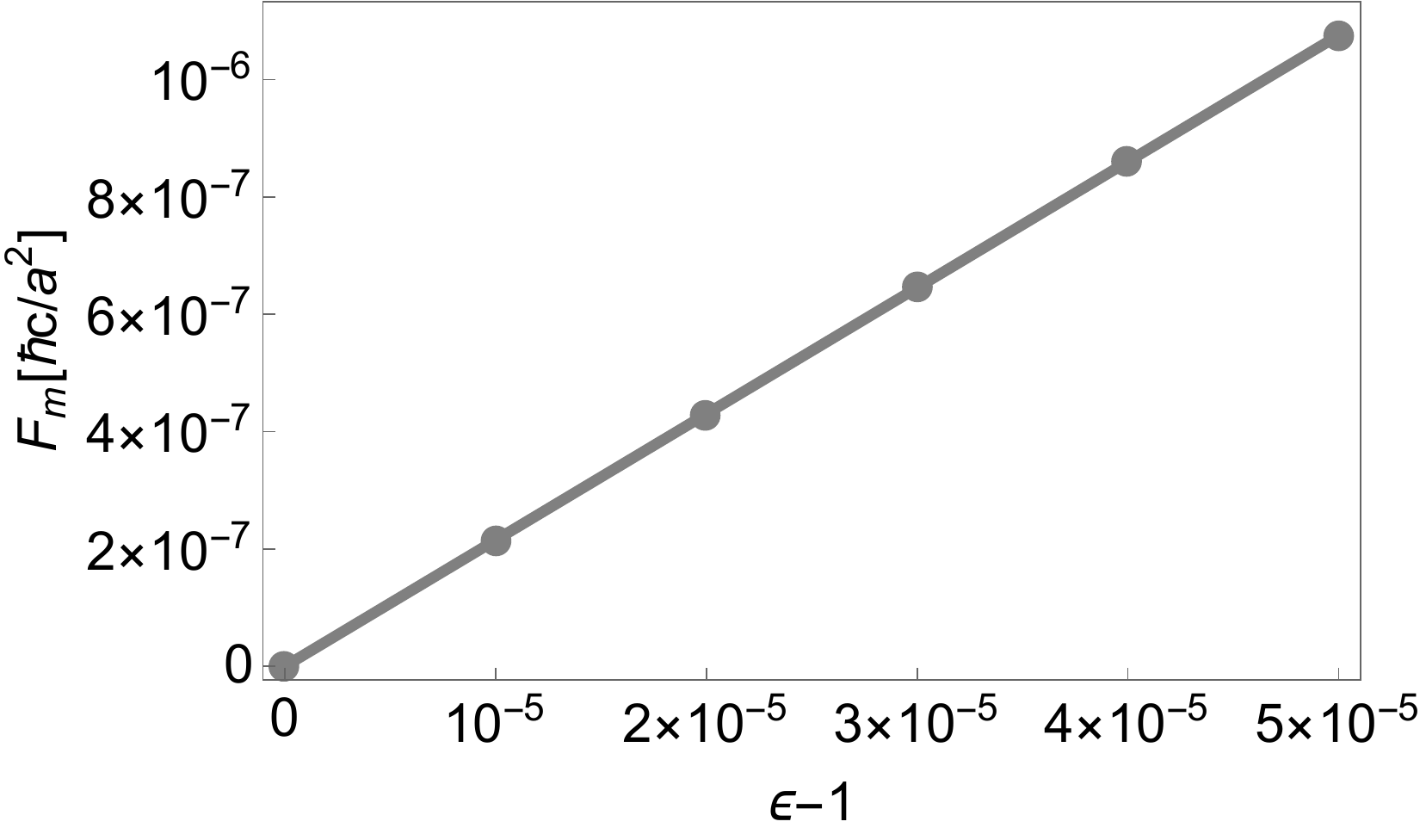}\centering
\caption {\small{$F_m$ in a dilute dielectric sphere inside vacuum: $\epsilon_1=\epsilon\approx1,\epsilon_2=\mu_{1,2}=1$ }}\label{dilute}
\end{figure}

To speed up the computation we used the asymptotic expansion of the Bessel functions. We carried the asymptotic expansion to sixth order to get an accurate result. The results are presented in Tab. \ref{table} and Figs.\ \ref{force} and \ref{dilute}.
\begin{table}[b]
\begin{ruledtabular}
\begin{tabular}{lccc}
& Our result & Result \cite{Milton1978} & Result \cite{Milton1997} \\ \hline 
Perfectly conducting spherical shell & $(0.0461\pm 0.0001)$ & $0.04617$ & -\\
$\epsilon_{1}=\mu_{2}=1$,\,\,\,\,\,\,\,\,\, $\epsilon_{2}=\mu_{1}=1.5$ & $(0.00162+0.00001)$ & - & $0.00155$\\
$\epsilon_{1}=\mu_{2}=2$,\,\,\,\,\,\,\,\,\,$\epsilon_{2}=\mu_{1}=1$ & $(0.00387\pm0.00001)$ & - & $0.0035$
\end{tabular}
\end{ruledtabular}
\caption{\small{Comparison between the macroscopic contribution to the force in units of $\hbar c a^{-2}$ obtained by Eq. (\ref{final force}) and published in Refs. \cite{Milton1978,Milton1997}. The results in Ref.~\cite{Milton1978} and Ref. \cite{Milton1997} are presented to the first disagreeing digit. The agreement with Ref.~\cite{Milton1978} is better since the calculation in Ref.~\cite{Milton1978} includes corrections to the first order approximation. In Ref.~\cite{Milton1997} only the first order was considered.}} 
\label{table}
\end{table}

Table \ref{table} shows that our method reproduces the limit of the perfectly conducting spherical shell (compared with Ref.~\cite{Milton1978}), and the case in which the speed of light is identical both inside and outside of the sphere (compared with Ref.~\cite{Milton1997}). For the latter, our results are more accurate than the ones of Ref.~\cite{Milton1997} as there only the first order in the asymptotic expansion was taken. However, while in Ref.~\cite{Milton1978} and \cite{Milton1997} the cutoff dependent terms are discarded by mathematical tricks, our calculation indicates that a cutoff dependent term, proportional to $1/ \delta$, does not vanish, meaning that the results are incomplete without knowledge of the microscopic structure.

Figure \ref{force} includes new results: it presents the macroscopic part of the force as a function of the dielectric constant in two cases: a dielectric sphere inside vacuum and a vacuum cavity inside a dielectric background.
The macroscopic part of the stress of a dielectric sphere inside vacuum is repulsive for low values of $\epsilon$, but for very large values of $\epsilon$ it becomes attractive. The case of a cavity is the opposite, the macroscopic part of the stress is attractive for low $\epsilon$ and becomes repulsive for high $\epsilon$ (but note that in the case of a dielectric sphere and the case of a vacuum cavity the force differs not just by a minus sign, but also in magnitude).

From these results one can estimate a correction to the surface tension $\gamma$ of a dielectric sphere. The typical macroscopic part of the Casimir force per unit area, as seen in Fig.\ \ref{force}, is:
\begin{equation}
\frac{F_m}{A}\sim\frac{0.01\hbar c}{4\pi a^{4}}.
\end{equation}
In a sphere $F / A = 2\gamma / a$ and therefore:
\begin{equation} \label{surface}
\gamma_m\sim \frac{10}{a[\text{nm}]^3}[\frac{\text{dyn}}{\text{cm}}].
\end{equation}
Note that the correction to the surface tension highly depends on the radius of the sphere: for large radii the surface tension is expected to equal its planar value, but at high curvatures the correction becomes significant. Comparing Eq. (\ref{surface}) to common materials such as the water-vapor interface (73 dyn/cm) and water-oil interface (57 dyn/cm) \cite{safran1994statistical}, we see that the correction starts being relevant only at sub-micron radii. 

One could experimentally test the Casimir contribution to the surface tension by acoustic resonances of micro droplets. The restoring force of deformations of the droplets is the capillary force due to surface tension \cite{LL6}. The eigenfrequencies thus depend on the value of the surface tension, including the Casimir contribution that is radius dependent. These eigenfrequencies appear as resonances with sound waves and therefore can be measured with precision. By making measurements with droplets of various radii one can identify the Casimir contribution.

Another feature shown in Fig.\ \ref{force} is that in the limit of $\epsilon \to \infty$, the sum over the two cases (vacuum-dielectric and dielectric-vacuum) gives the famous result of the stress in a perfectly conducting spherical shell (see Fig.~\ref{sum} in the introduction). That is due to the fact that when the boundary is a perfect reflector, the inside of the sphere ``does not know" about the outside and {\it vice versa}. Therefore the stress inside a perfectly reflecting spherical shell is the same as the stress of a vacuum cavity inside a high $\epsilon$ background, and the stress outside a perfectly reflecting spherical shell is the same as the stress of vacuum surrounding a high $\epsilon$ sphere.

Figure \ref{dilute} presents the most surprising of our results: it shows that the Casimir force of a very dilute homogeneous sphere inside vacuum, when $\epsilon-1 \approx 1$, is to first order linear in $\epsilon-1$ rather then quadratic. This is a feature of the finite macroscopic term alone (the cutoff dependent terms are to first order quadratic in $\epsilon-1$).

This result is a contradiction to the present understanding that the Casimir force in dilute systems can be derived by pairwise summation of van der Waals interactions using the Casimir-Polder retarded potential \cite{Casimir1948}. The Casimir-Polder potential between two identical atoms is proportional to $\alpha^2$, where $\alpha$ is the atoms' polarizability. In the dilute case $\epsilon\approx 1+4 \pi N \alpha$ and therefore pairwise summation would give an interaction proportional to $(\epsilon-1)^2$. The Casimir force between two dilute planar walls is indeed quadratic in $\epsilon-1$ and coincides with the force derived by pairwise summation. However, that is not the case in a sphere.

A physical example of a very dilute sphere is a Bose--Einstein condensate of alkali atoms \cite{BEC}. Using light one can, in principle, create sharp boundaries (limited of course by the diffraction of light) and confine the atomic gas if not in a sphere, but certainly in the related case of a cylinder. Interestingly, a linear dependence on the density was also seen \cite{Matzliah} for the optomechanical strain in a cold cloud of atoms. There a plane wave of light, illuminating the atomic cloud, is focused by the lens-like shape of the cloud. In turn, the atoms experience the recoil of the light and expand when released. Naively one would expect that both the focusing and the recoil is linear in the density of the atoms such that the overall effect is quadratic in density, but experiment \cite{Matzliah} proves it is not --- it is linear. Similarly, we have found theoretically that in the dilute limit the Casimir force is linear in density. The experiment \cite{Matzliah} also shows that a dilute gas with peak $\varepsilon = 0.00002$ still acts as a dielectric, and not as a collection of individual atoms. Therefore, our model of the dilute dielectric sphere may remain valid even in the extremely dilute limit.

\begin{figure}[!tbp]
  \centering
{\includegraphics[width=0.45\textwidth]{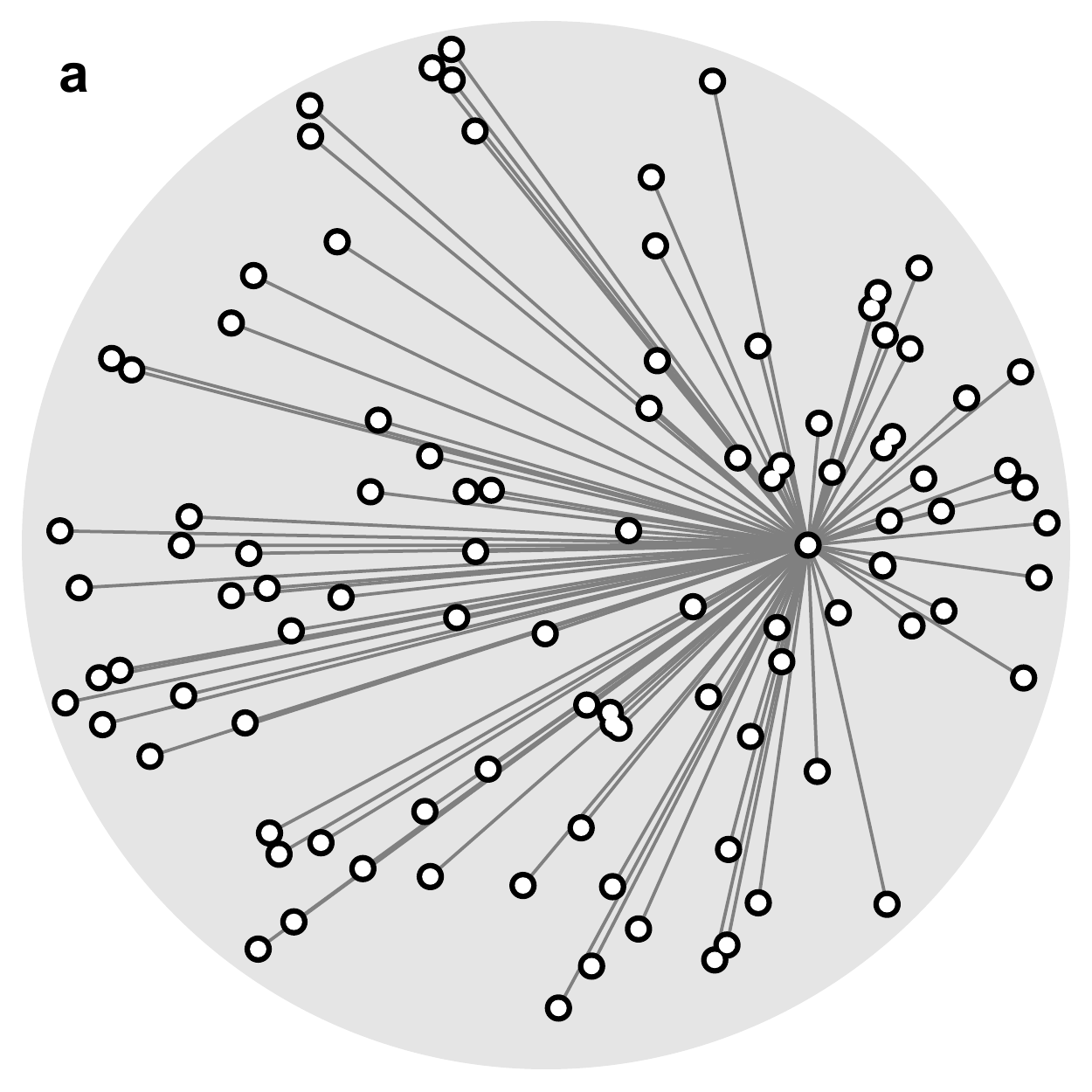}\label{fig:f1}}
  \hfill
{\includegraphics[width=0.45\textwidth]{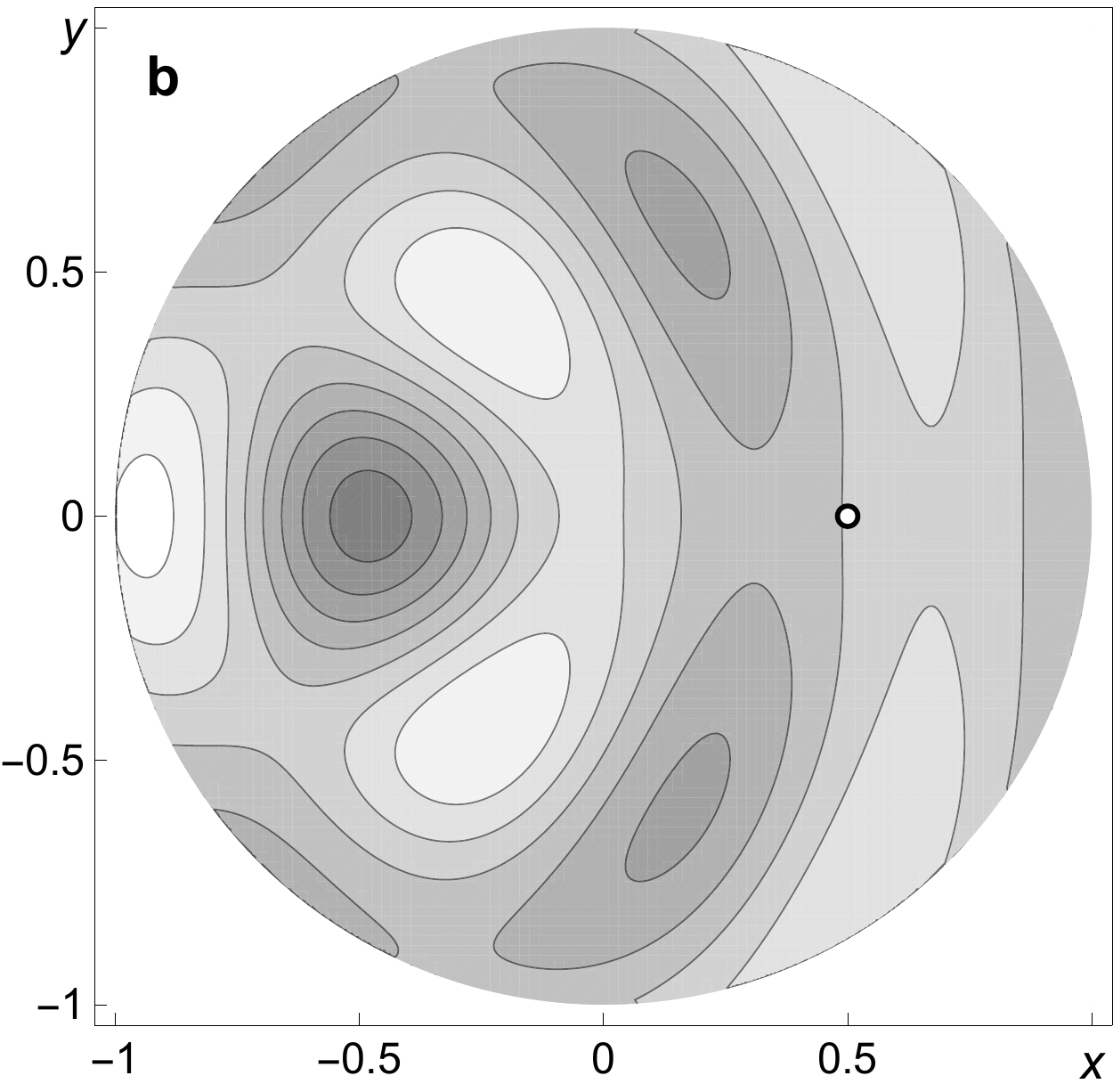}\label{fig:f2}}
\caption {\small{Direct interactions versus scattering. a: diagram illustrating the physical picture of the Casimir force as the result of direct molecular interactions. In the dilute limit, a given molecule (white point) interacts with each of the other molecules (other white points) by Casimir-Polder forces; triple or higher interactions can be neglected. In this picture the Casimir stress would depend quadratically on the density, {\it i.e.} quadratically on $\epsilon-1$. b: diagram showing the scattering of a wave emitted by the molecule (white point) at the boundary of the medium. The reflected wave gives rise to the Casimir stress. As one reflection already generates stress, the Casimir stress should scale linearly with $\epsilon-1$ in the dilute limit, which does agree with our numerical results. The figure shows the imaginary part of the difference between a wave in the medium and the outgoing wave (for a scalar wave in 2D, for simplicity).} }  \label{reflections}
\end{figure}

Our results contradict the accepted picture of the Casimir forces, but they lend support to an alternative picture (Fig.~\ref{reflections}) that follows naturally from Lifshitz theory. As Sec.~II describes, the starting point of Lifshitz theory is the fluctuation-dissipation theorem of Eq.~(\ref{theorem}) that connects the noise of the field as described in the field correlator to the classical Green function emitted at source point $\bm{r}'$ with unity strength and received at point $\bm{r}'$. The reflected part of the Green function gives rise to the renormalized stress. For piece-wise homogenous materials such as dielectric plates \cite{Casimir1948a,LifshitzE.M.1956,Dzyaloshinskii1961} or the dielectric sphere considered here, the reflections occur at the interfaces between the different homogeneous regions. In the dilute limit, each reflection goes with a factor of $\epsilon-1$. When the boundaries are flat, the first reflection gets canceled in the force calculation. This is due to a cancellation between the magnetic and electric contributions to the stress, which is attributed to the fact that in a single reflection the magnetic field gains a minus sign ($\pi$ phase) relative to the electric field. This means that the lowest-order contribution goes with $(\epsilon-1)^2$ as in the pairwise summation of van der Waals forces. However, this symmetry does not apply when the walls are curved, as seen by our results. The first reflection makes a non-zero contribution to the Casimir force, which is why we get a contribution linear in $\epsilon-1$ in a sphere. The dielectric sphere is thus the first crucial case known so far for discriminating between Lifshitz theory and pairwise van der Waals theory. 

The fluctuation-dissipation theorem, Eq.~(\ref{theorem}), justifies the picture of the Casimir forces as being caused by reflections of virtual waves emitted and received at each molecule of the medium, the reflections being caused by all the other molecules acting as a medium. The case of the sphere rules out the picture of the Casimir force as a multitude of pair-wise van der Waals interactions between molecules. Yet in Ref.\ \cite{Brevik1999} an equivalence was shown between the Casimir force and the pairwise summation in a dilute sphere.  Note that this calculation, though impressive, mistakenly changes the order of limits (the limit $r\to a$ is taken before the $l$ summation) which is what causes the linear term to vanish. The mistake is similar to changing the order between the limit and the integral of the expression $\lim_{\varepsilon\to0}\int\limits _{0}^{\infty}\varepsilon e^{-r\varepsilon} dr$, which changes the result from 1 to 0. By taking the correct order of limits, one obtains the linear contribution. This means that the pairwise interaction are a part of, but not the entire story.

\section{The inclusion of dispersion} \label{dispersion}
All the above results do not include dispersion in them, {\it i.e.} they are an idealization of materials whose response to fluctuations does not depend on the frequency of the fluctuation, which is of course not physical. When trying to account for dispersion, for instance by using a simple model for the dielectric constant:
\begin{equation}
\epsilon=1+\frac{\omega_{P}^{2}}{\omega_{0}^{2}-\omega^{2}-i\gamma\omega},
\end{equation}
we get that the terms $\delta^{-3}$, $\delta^{-2}$ and $\delta$ vanish from the expansion in Eq. (\ref{fit2}) such that the highest order term is the logarithm. As we increase $\omega_p$ and $\omega_0$, $F_m$ becomes very large. That is even though the limit of $\omega_p$,$\omega_0 \to \infty$ is suppose to coincide with the dispersionless case. We suspect that there are two reasons to it: 1. The dispersion introduces a new scale into the system, and so the argument of taking $a$ to be the scale of the logarithm, because it is the only scale, does not apply anymore. 2. The clear distinction between the finite macroscopic contribution, and the cutoff dependent terms which are related to the microscopic description is lost when dispersion is included. The terms are mixed within each other, and although the total force must coincide between the dispersionless case and the limit of $\omega_p$,$\omega_0 \to \infty$, $F_m$ does not coincide since it assumes separation between the different contributions, which does not exist with dispersion.
Our conclusions from this discussion are as follows: the method we introduced to calculate the macroscopic contribution to the Casimir self-stress of a sphere applies only to non-dispersive materials. It can provide us with insights and estimations of orders of magnitude, but for accurate calculations of real materials, one can not separate the macroscopic contribution: a microscopic calculation is needed either way.

\section{Summary}
In this paper we proposed a method to calculate the Casimir stress in a homogeneous sphere inside a homogeneous background for dispersionless materials, in which the distance from the sphere plays the role of a regularizer. We believe that our procedure is more accurate than previous calculations, as the mathematical operations and order of limits we take have a mathematical justification. We give an estimation for the correction to the surface tension that comes from the macroscopic effect in dispersionless materials. The correction has a special dependence on the radius: it is proportional to $a^{-3}$, which as expected, makes it noticeable only at very small radii. However, one could probably measure the effect using acoustic resonances with micro droplets.

Our findings agree with previous results in several cases, including the famous limit of a perfectly conducting spherical shell, but disagrees in other cases. The most important disagreement is the Casimir self-stress of a very dilute dielectric ball, for which our results shutter the picture of the equivalence between the macroscopic effect and pairwise summation, by showing the existence of an additional term that cannot be explained by pairwise summation. We give an interpretation for this term using a picture of reflections of waves from the boundaries. While the Casimir force remains ``one of the least intuitive consequences of quantum electrodynamics'' \cite{VanEnk1995}, we have uncovered evidence for a physical picture that may eventually lift part of the mystery. 

\begin{acknowledgments}
We would like to thank David Andelman, Efi Efrati, Itay Griniasty, Moti Milgrom, Jonathan Drori and Yuval Rosenberg for stimulating discussions. Our work was supported by the European Research Council and the Israel Science Foundation, a research grant from Mr. and Mrs. Louis Rosenmayer and from Mr. and Mrs. James Nathan, and the Murray B. Koffler Professorial Chair.
\end{acknowledgments}
\appendix

\section{The calculation of the stress components} \label{app A}
In this appendix we derive Eqs.\ (\ref{diag}-\ref{coefficients}). Applying the machinery of Lifshitz theory we calculate the relevant Green's function. More precisely, we solve Eq.\ (\ref{green function eq}) with the appropriate boundary conditions: continuity of
\[
\hat{r}\times G\,,\,\,\,\hat{r}\cdot\epsilon G\,,\,\,\,\hat{r}\times\frac{1}{\mu}\vec{\nabla}\times G\,,\,\,\,\hat{r}\cdot\vec{\nabla}\times G
\]
and the requirement that $G$ is finite at $r\to0$ and behaves like a spherical outgoing wave at $r\to\infty$.
Following Ref. \cite{Milton1978} where this calculation was done for a perfectly conducting spherical shell, and generalizing it to dielectrics we obtain the solution:
\begin{equation}
\begin{gathered}G\left(\bm{r},\bm{r}',\omega\right)=\sum_{lm}\{\mu F_{l}\left(r,r'\right)\bm{X}_{lm}\left(\theta,\phi\right)\otimes\bm{X}_{lm}^{\ast}\left(\theta',\phi'\right)\\
\,\,\,\,\,\,\,\,\,\,\,\,\,\,\,\,\,\,\,\,\,\,\,\,\,\,\,\,\,\,\,\,\,\,\,\,\,\,\,\,\,\,\,\,\,\,\,+\frac{c^{2}}{\epsilon\omega^{2}}\vec{\nabla}\times\left[G_{l}\left(r,r'\right)\bm{X}_{lm}\left(\theta,\phi\right)\otimes\bm{X}_{lm}^{\ast}\left(\theta',\phi'\right)\right]\times\overleftarrow{\nabla}'\}
\end{gathered}
\end{equation}
where the \(\bm{X}_{lm}\) are the vector spherical harmonics related to the well-known spherical harmonics \(Y_{lm}\) as
\begin{equation}
\bm{X}_{lm}\left(\theta,\phi\right)=\frac{1}{\left[l\left(l+1\right)\right]^{\frac{1}{2}}}\frac{1}{i}\left(\bm{r}\times\vec{\nabla}\right)Y_{lm}\left(\theta,\phi\right).
\end{equation}
The \(F_l\) and \(G_l\) are scalar functions representing the two different polarizations defined by
\begin{equation}
\begin{gathered}F_{l}=\begin{cases}
ik_{1}j_{l}\left(k_{1}r_{<}\right)\left(A_{\mu}^{l}j_{l}\left(k_{1}r_{>}\right)+h_{l}\left(k_{1}r_{>}\right)\right) & r,r'<a\\
ik_{2}h_{l}\left(k_{2}r_{>}\right)\left(B_{\mu}^{l}h_{l}\left(k_{2}r_{<}\right)+j_{l}\left(k_{2}r_{<}\right)\right) & r,r'>a
\end{cases}\\
G_{l}=\begin{cases}
ik_{1}j_{l}\left(k_{1}r_{<}\right)\left(A_{\epsilon}^{l}j_{l}\left(k_{1}r_{>}\right)+h_{l}\left(k_{1}r_{>}\right)\right) & r,r'<a\\
ik_{2}h_{l}\left(k_{2}r_{>}\right)\left(B_{\epsilon}^{l}h_{l}\left(k_{2}r_{<}\right)+j_{l}\left(k_{2}r_{<}\right)\right) & r,r'>a
\end{cases}\\
A_{\nu}^{l}=\frac{\nu_{1}h_{l}\left(k_{1}a\right)\partial_{r}\left(rh_{l}\left(k_{2}r\right)\right)-\nu_{2}h_{l}\left(k_{2}a\right)\partial_{r}\left(rh_{l}\left(k_{1}r\right)\right)}{\nu_{2}h_{l}\left(k_{2}a\right)\partial_{r}\left(rj_{l}\left(k_{1}r\right)\right)-\nu_{1}j_{l}\left(k_{1}a\right)\partial_{r}\left(rh_{l}\left(k_{2}r\right)\right)}|_{r=a}\\
B_{\nu}^{l}=\frac{\nu_{1}j_{l}\left(k_{1}a\right)\partial_{r}\left(rj_{l}\left(k_{2}r\right)\right)-\nu_{2}j_{l}\left(k_{2}a\right)\partial_{r}\left(rj_{l}\left(k_{1}r\right)\right)}{\nu_{2}h_{l}\left(k_{2}a\right)\partial_{r}\left(rj_{l}\left(k_{1}r\right)\right)-\nu_{1}j_{l}\left(k_{1}a\right)\partial_{r}\left(rh_{l}\left(k_{2}r\right)\right)}|_{r=a}\\
\nu=\epsilon,\mu\,\,\,\,,\,\,\,\,k_{i} = \sqrt{\epsilon_{i}\mu_{i}}\frac{\omega}{c}
\end{gathered}
\end{equation}
where $j_l$ and $h_l$ are spherical Bessel and spherical Hankel functions, and \(r_>\) (\(r_<\)) is the larger (smaller) between \(r\) and \(r'\).

In accordance with the standard Lifshitz renormalization, we subtract the free Green's function responsible for the direct interaction, which is equivalent to the omission of
\(
ik_{1}j_{l}\left(k_{1}r_{<}\right)h_{l}\left(k_{1}r_{>}\right)
\)
and
\(
ik_{2}j_{l}\left(k_{2}r_{<}\right)h_{l}\left(k_{2}r_{>}\right)
\) from the scalar Green's function inside and outside, respectively. We are then left with
\begin{eqnarray}
F_{l}^{\text{ren}} & = & i\begin{cases}
A_{\mu}^{l}k_{1}j_{l}\left(k_{1}r\right)j_{l}\left(k_{1}r'\right) & r,r'<a\\
B_{\mu}^{l}k_{2}h_{l}\left(k_{2}r\right)h_{l}\left(k_{2}r'\right) & r,r'>a
\end{cases}\\
G_{l}^{\text{ren}} & = & i\begin{cases}
A_{\epsilon}^{l}k_{1}j_{l}\left(k_{1}r\right)j_{l}\left(k_{1}r'\right) & r,r'<a\\
B_{\epsilon}^{l}k_{2}h_{l}\left(k_{2}r\right)h_{l}\left(k_{2}r'\right) & r,r'>a.
\end{cases} \nonumber
\end{eqnarray}
Next we use the renormalized Green's function to calculate the correlators in Eq.\ (\ref{correlatorsim}). There we need to write our expressions for imaginary frequencies. In order to simplify the functions we use the relations between the spherical Bessel and Hankel functions and the modified Bessel functions \(I\) and \(K\):
\begin{equation}
j_{l}\left(ix\right)=\sqrt{\frac{\pi}{2x}}i^{l}I_{l+\frac{1}{2}}\left(x\right),\quad
h_{l}\left(ix\right)=-\sqrt{\frac{2}{\pi x}}i^{l}\left(\left(-1\right)^{l}K_{l+\frac{1}{2}}\left(x\right)\right).
\end{equation}
The scalar Green's functions then take the form
\begin{eqnarray} \label{scalar Green's functions renormalized}
F_{l}^{\text{ren}} & = & \begin{cases}
\tilde{A}_{\mu}^{l}\frac{I_{l+\frac{1}{2}}\left(\kappa_{1}r\right)I_{l+\frac{1}{2}}\left(\kappa_{1}r'\right)}{\sqrt{rr'}} & r,r'<a\\
\tilde{B}_{\mu}^{l}\frac{K_{l+\frac{1}{2}}\left(\kappa_{2}r\right)K_{l+\frac{1}{2}}\left(\kappa_{2}r'\right)}{\sqrt{rr'}} & r,r'>a
\end{cases}\\
G_{l}^{\text{ren}} & = & \begin{cases}
\tilde{A}_{\epsilon}^{l}\frac{I_{l+\frac{1}{2}}\left(\kappa_{1}r\right)I_{l+\frac{1}{2}}\left(\kappa_{1}r'\right)}{\sqrt{rr'}} & r,r'<a\\
\tilde{B}_{\epsilon}^{l}\frac{K_{l+\frac{1}{2}}\left(\kappa_{2}r\right)K_{l+\frac{1}{2}}\left(\kappa_{2}r'\right)}{\sqrt{rr'}} & r,r'>a
\end{cases} \nonumber
\end{eqnarray}
where \(\tilde{A}_{\nu}^{l}=\frac{\pi}{2}\left(-1\right)^{l}A_{\nu}^{l}\,\,\,,\,\,\,\tilde{B}_{\nu}^{l}=\frac{2}{\pi}\left(-1\right)^{l}B_{\nu}^{l}\) and \(\kappa_{i} = \sqrt{\epsilon_{i}\mu_{i}}\frac{\xi}{c}\).

We combine Eqs.\ (\ref{stress}) and (\ref{correlatorsim}) to compute the stress, taking the limit $\bm{r} \to \bm{r}'$. We use the dimensionless variables: $y=a\kappa$~,~$y_{1,2}=y\sqrt{\epsilon_{1,2}\mu_{1,2}}$, $\rho={r}/{a}$, redefine the scalar Green's functions (such that they are dimensionless as well): $\tilde{F}_{l}=aF_{l}\,\,\,,\,\,\,\tilde{G}_{l}=aG_{l}$, and obtain:
\begin{equation} \label{stressr}
\sigma_{r}^{r}\left(\bm{r}\right)=\frac{\hbar c}{8\pi^{2}r^{2}a^{2}}\sum_{l}\int\limits _{0}^{\infty}\left(2l+1\right)\left(l\left(l+1\right)+y_{i}^{2}\rho^{2}-\frac{\text{d}}{\text{d}\rho}\rho\frac{\text{d}}{\text{d}\rho'}\rho'\right)\left(\tilde{F}_{l}+\tilde{G}_{l}\right)|_{\rho'\to\rho}\,\text{d}y
\end{equation}
\begin{equation} \label{stresstheta}
\sigma_{\theta}^{\theta}\left(\bm{r}\right)=\sigma_{\phi}^{\phi}\left(\bm{r}\right)=-\frac{\hbar c}{8\pi^{2}r^{2}a^{2}}\sum_{l}\int\limits _{0}^{\infty}\left(2l+1\right)l\left(l+1\right)\left(\tilde{F}_{l}+\tilde{G}_{l}\right)|_{\rho'\to\rho}\,\text{d}y
\end{equation}
\begin{equation}
\sigma_{m}^{n}=0\,\,\,\text{for}\,\,\,n\neq m
\end{equation}
where the index $i$ is 1 if the function is evaluated inside the sphere and 2 if it is evaluated outside the sphere.

For simplicity we define: $g_{l}\left(x\right)=\tilde{F}_{l}\left(x\right)+\tilde{G}_{l}\left(x\right)$, $C_{\text{i}}\left(x\right)=\tilde{A}_{\epsilon}^{l}\left(x\right)+\tilde{A}_{\mu}^{l}\left(x\right)$, $C_{\text{o}}\left(x\right)=\tilde{B}_{\epsilon}^{l}\left(x\right)+\tilde{B}_{\mu}^{l}\left(x\right)$ and finally arrive at Eqs.\ (\ref{diag}-\ref{coefficients}).

\section{The behavior of $\Delta r^2\sigma_{r}^{r}(\delta)$} \label{app B}

To analyze the behavior of $\Delta r^2\sigma_{r}^{r}(\delta)$ at $\delta\ll 1$ we apply the asymptotic expansion of the modified Bessel functions \cite{Bickley1966}:
\begin{eqnarray} \label{asymptotics}
I_{\nu}\left(\nu z\right) & = & \frac{e^{\nu\eta}}{\left(2\pi\nu\right)^{\frac{1}{2}}\left(1+z^{2}\right)^{\frac{1}{4}}}\sum_{k=0}^{\infty}\frac{U_{k}\left(p\right)}{\nu^{k}}\\
K_{\nu}\left(\nu z\right) & = & \left(\frac{\pi}{2\nu}\right)^{\frac{1}{2}}\frac{e^{-\nu\eta}}{\left(1+z^{2}\right)^{\frac{1}{4}}}\sum_{k=0}^{\infty}\left(-1\right)^{k}\frac{U_{k}\left(p\right)}{\nu^{k}} \nonumber
\end{eqnarray}
where
\begin{equation}
\eta = \left(1+z^{2}\right)^{\frac{1}{2}}+\ln\left(\frac{z}{1+\left(1+z^{2}\right)^{\frac{1}{2}}}\right),\quad
p = \left(1+z\right)^{-\frac{1}{2}} \nonumber
\end{equation}
and the $U_{k}(p)$'s are polynomials defined by the recurrence relation:
\begin{eqnarray}
U_{k+1}\left(p\right) & = & \frac{1}{2}p^{2}\left(1-p^{2}\right)U{}_{k}^{'}\left(p\right)+\frac{1}{8}\int\limits _{0}^{p}\left(1-5t^{2}\right)U_{k}\left(t\right)dt \,, \nonumber \\
U_{0}\left(p\right) & = & 1 \,.
\end{eqnarray}
When $\delta=0$ the function $\Delta r^2 \sigma_{r}^{r}(0)$ diverges. When $\delta\neq0$ we analyze $\Delta r^2 \sigma_{r}^{r}(\delta)$ by making the change of variables:
\begin{equation}
r=\sqrt{\epsilon_{i}\mu_{i}x^{2}+\nu^{2}},\quad
t=\sin^{-1}\left(\frac{\nu}{\sqrt{\epsilon_{i}\mu_{i}}x}\right)
\end{equation}
and, in order to get a qualitative behavior, we replace the sum over $l$ by an integral, and disregard the contribution of small $x$'s and $l$'s. We get (for $\delta\ll 1$):
\begin{equation}
\sigma_{r}^{r} \sim \sum_{n=0}^{\infty}\sum_{m=0}^{\infty}C_{nm}\int\limits _{r_{0}}^{\infty}e^{-2r\delta}r^{2-n}\delta^{m}\text{d}r.
\end{equation}
The result is a sum over integrals, where each of the integrals can be expanded for small $\delta$ by a Laurent series plus a Taylor series multiplied by a logarithm. The highest order term in the Laurent series is $\delta ^{-3}$, which is obtained for $n=m=0$. The highest order term in the Taylor-times-logarithm series is $\log(\delta)$ and it is obtained for $n=3, m=0$. Hence we get the series in Eq.\ (\ref{expansion}).

\section{Convergence of the $\Delta r^2 \sigma_{r}^{r}(\delta)$ fit} \label{app C}
\begin{figure}[!tbp]
  \centering
  {\includegraphics[width=0.45\textwidth]{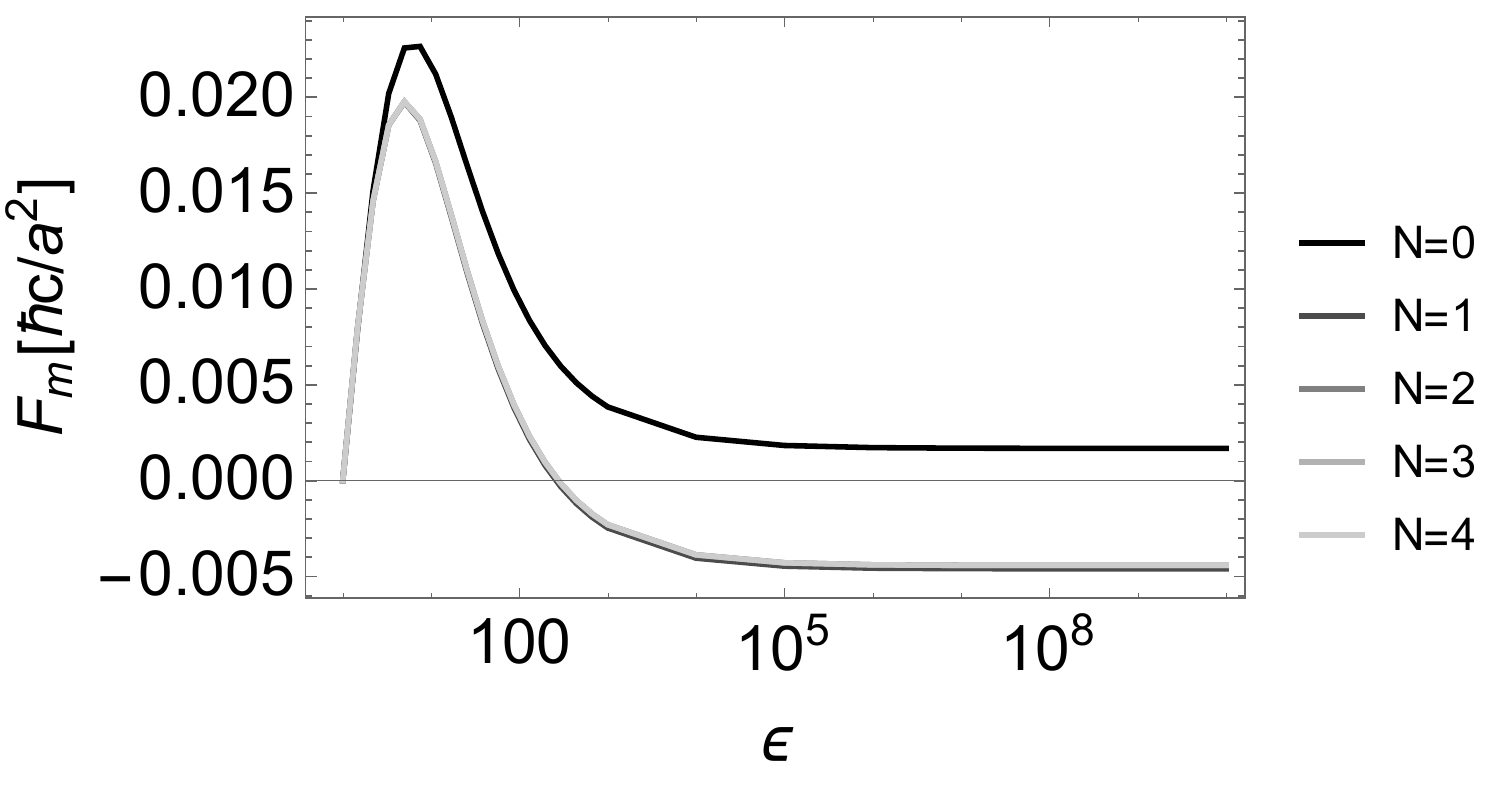}\label{convergence}}
  \hfill
  {\includegraphics[width=0.45\textwidth]{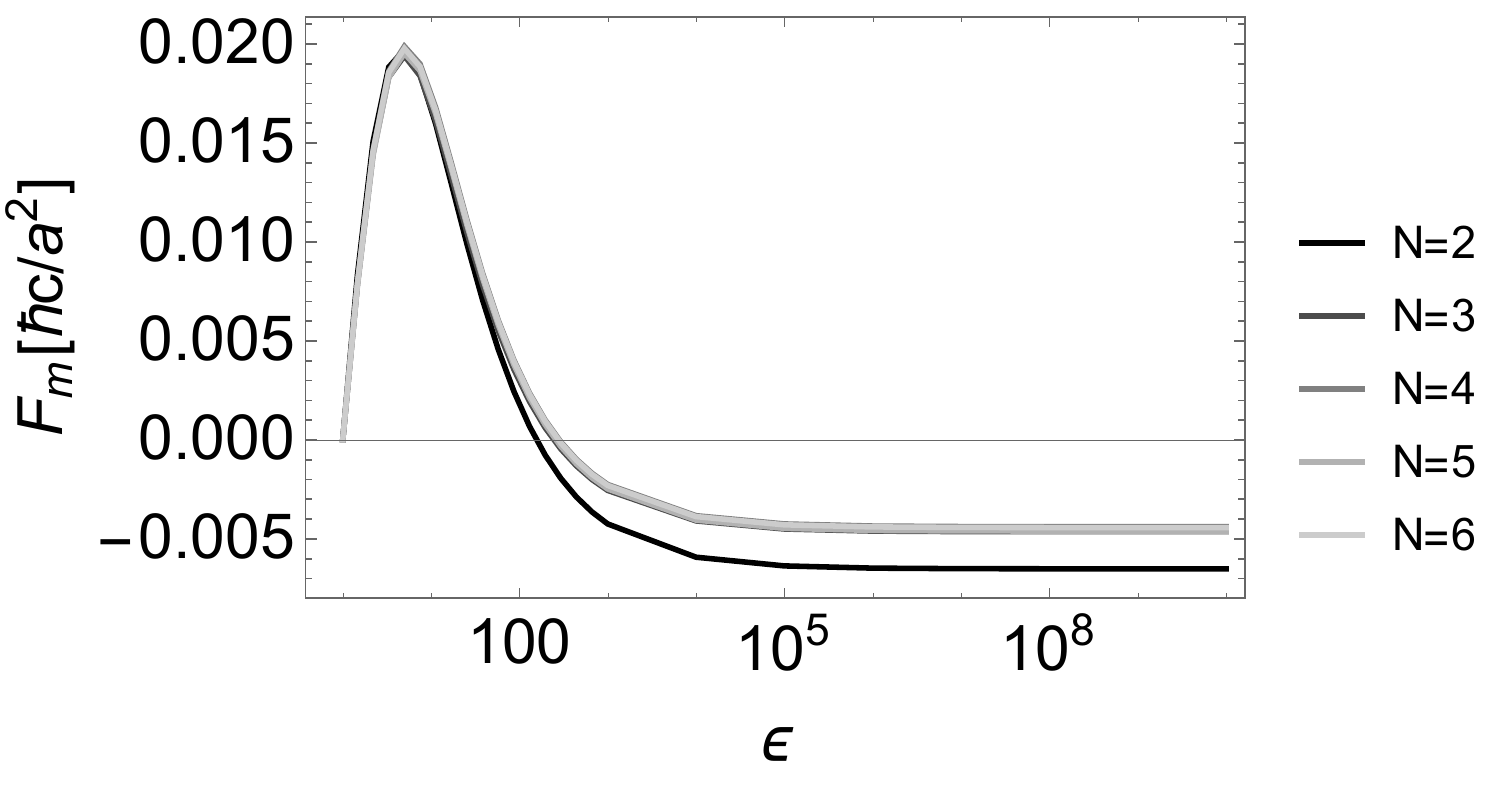}\label{convergence2}}
  \caption{\small{
  Convergence of the fit ({\bf a}) and of the asymptotic expansion ({\bf b}). The figures show
  $F_m$ with the parameters $\epsilon_2=\mu_{1,2}=1$, $\epsilon_1=\epsilon$ as a function of $\epsilon$. In ({\bf a})  $N$ is the highest order taken in the linear fit: $y=\sum_{n=-3}^{N}a_{n}^{\mbox{i}}\delta^{n}+\sum_{n=0}^{N}b_{n}^{\mbox{\ensuremath{\text{i}}}}\delta^{n}\log\left(\delta\right)$ while in ({\bf b})  $N$ is the highest order taken in the asymptotic expansion.}}
\end{figure}

The series in Eq.\ (\ref{expansion}) is infinite. However, when fitting a function numerically to a model, the model must have a finite number of parameters. Luckily, as we approach the boundary, all the positive powers of $\delta$ in the expansion approach zero, and therefore become more and more negligible. Thus, in order to extract the constant to good accuracy, we can fit the numerically evaluated function to a truncated series while making sure that we include all the elements in the expansion which are not negligible. To achieve this, we fit the function several times, each time adding another higher order term, and stop when the constant converged within the desired accuracy. Figure \ref{convergence} shows the convergence process and the reason for choosing $N=4$ as the maximal order in our calculations [Eq. (\ref{fit2})].

We used a similar procedure to determine the number of orders in the asymptotic expansion we take into account. We increased the number of orders until we reached a converging result, as depicted in Fig. \ref{convergence2}.

\begin{figure}[h!]
\includegraphics[width=8cm]{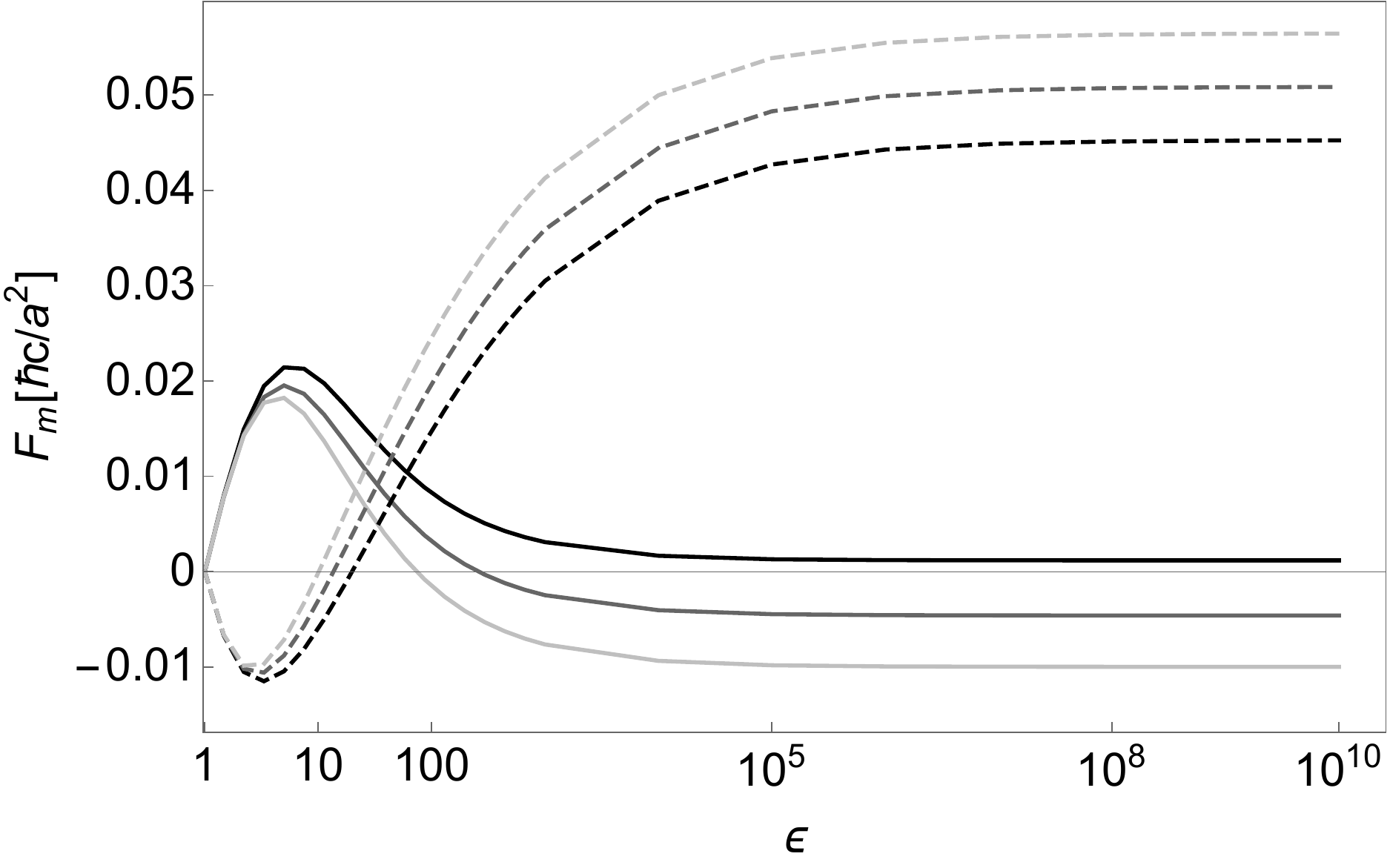}\centering
\caption {\small{$F_m$ as a function of $\epsilon$ with the parameter fit of Eq. (\ref{fit2}), $\delta$, replaced by $\alpha \delta$. From bright to dark: $\alpha=1/2, 1, 2$. Full line: $\epsilon_2=\mu_{1,2}=1$, $\epsilon_1=\epsilon$. Dashed line: $\epsilon_1=\mu_{1,2}=1$, $\epsilon_2=\epsilon$.}}\label{log}
\end{figure}

Figure \ref{log} shows  the modification of the results if we replace $\delta$ by $\alpha \delta$ where $\alpha$ is of order unity. We see that that for small dielectric constants the results barely change, whereas for large dielectric constants the force does depend on $\alpha$ and can even change from attractive to repulsive.

\end{document}